\documentclass[12pt,dvips,final]{article}

\usepackage{times}		% Default font: Times Roman, 12pt
\usepackage{graphicx}		% for inclusion of EPS figures
\usepackage{color}		% for colored text and boxes
\usepackage{fancyheadings_b}	% my modification of 'fancyheadings' package
\usepackage{wrapfig}		% for embedding figures/tables within text
\usepackage{bm}			% Bold math
\usepackage{natbib}		% References
\usepackage[english]{babel}
\newcommand{\etal}      {\mbox{et al.\ }}%
\newcommand{\eg}        {\mbox{e.g.}}%
\newcommand{\tsim}      {\ensuremath{\sim}}%
\newcommand{\Lya}       {\ensuremath{\mbox{\textrm{Ly}}\alpha}}%
\newcommand{\Ha}        {\ensuremath{\mbox{\textrm{H}}\alpha}}%
\newcommand{\Hb}        {\ensuremath{\mbox{\textrm{H}}\beta}}%
\newcounter{ion}
\newcommand{\ion}[2]	{\setcounter{ion}{#2}\mbox{#1\,{\small\@\Roman{ion}}}}%
\newcommand{\HII}       {\mbox{\ion{H}{2}}}%
\newcommand{\OI}        {\mbox{[\ion{O}{1}]}}%
\newcommand{\OII}       {\mbox{[\ion{O}{2}]}}%
\newcommand{\OIII}      {\mbox{[\ion{O}{3}]}}%
\newcommand{\SII}       {\mbox{[\ion{S}{2}]}}%
\newcommand{\SIII}      {\mbox{[\ion{S}{3}]}}%
\newcommand{\lam}       {\ensuremath{\lambda}}%
\newcommand{\lamlam}    {\ensuremath{\lambda\lambda}}%
\newcommand{\farcs}     {\mbox{\ensuremath{.\mkern-5mu^{\prime\prime}}}}%
\newcommand{\lesssim} {\mbox{\rlap{\hbox{\lower3pt\hbox{\ensuremath{\sim}}}}\raise2pt\hbox{\ensuremath{<}}}}%
\newcommand{\gtrsim}  {\mbox{\rlap{\hbox{\lower3pt\hbox{\ensuremath{\sim}}}}\raise2pt\hbox{\ensuremath{>}}}}%
\newcommand{\HST}       {\emph{HST}}%
\newcommand{\Spitzer}   {\emph{Spitzer}}%
\newcommand{\GALEX}     {\emph{GALEX}}%
\newcommand{\SINGS}     {\emph{SINGS}}%
\definecolor{navy}{rgb}{0,0,0.63}
\definecolor{lblue}{rgb}{0.2,0.8,1}
\definecolor{lightcyan}{rgb}{0.5,1,1}
\definecolor{dgreen}{rgb}{0.0,0.5,0.0}
\definecolor{lgreen}{rgb}{0.56,0.93,0.49}
\definecolor{gold}{rgb}{0.90,0.75,0.00}
\definecolor{orange}{rgb}{1.0,0.5,0.0}
\definecolor{maroon}{rgb}{0.6,0,0.2}
\definecolor{grey}{rgb}{0.1,0.1,0.1}

\newlength{\txw}
\newlength{\txh}

\begin{document}
\addtocounter{page}{-2}

%%%%%%%%%%%%%%%%%%%%%%%%%%%%%%%%%%%%%%%%%%%%%%%%%%%%%%%%%%%%%%%%%%%%%%%%%%%%
%        H E R E    W E    S T A R T    T H E    C O V E R    P A G E
%%%%%%%%%%%%%%%%%%%%%%%%%%%%%%%%%%%%%%%%%%%%%%%%%%%%%%%%%%%%%%%%%%%%%%%%%%%%

%--- Begin of page format definitions...
\pagestyle{empty}
\setlength{\textwidth}{8.5in}
\setlength{\textheight}{11.0in}
\setlength{\oddsidemargin}{-1.0in}
\setlength{\evensidemargin}{\oddsidemargin}
\setlength{\parindent}{0pt}
%--- End of page format.

\noindent\hspace*{0.0in}\begin{minipage}[t]{8.5in}
   \vspace*{-1.35in}\includegraphics[width=8.5in]{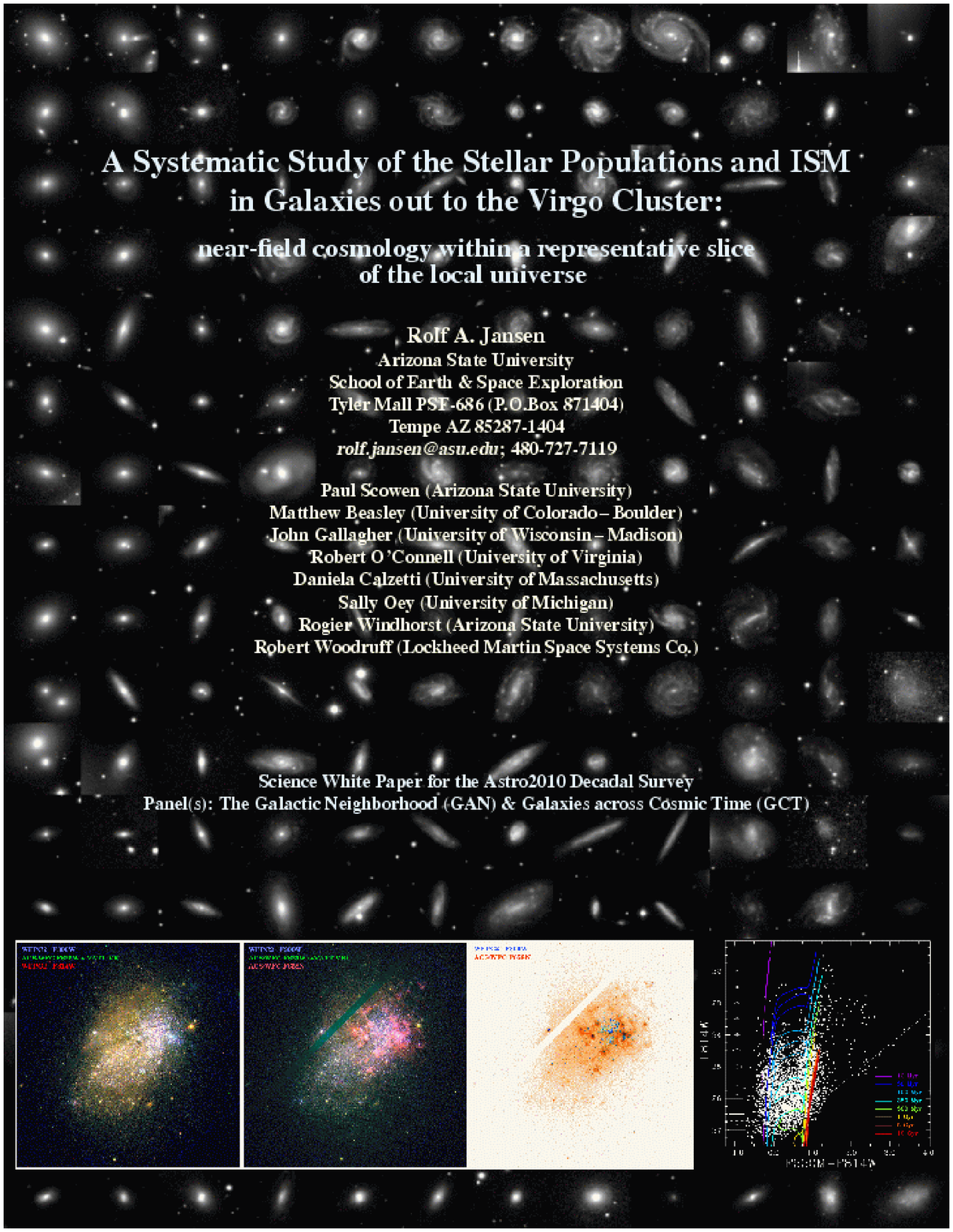}
\end{minipage}
\newpage

%%%%%%%%%%%%%%%%%%%%%%%%%%%%%%%%%%%%%%%%%%%%%%%%%%%%%%%%%%%%%%%%%%%%%%%%%%%%
%              H E R E    W E    S T A R T    T H E    T E X T
%%%%%%%%%%%%%%%%%%%%%%%%%%%%%%%%%%%%%%%%%%%%%%%%%%%%%%%%%%%%%%%%%%%%%%%%%%%%

%--- Begin of page format definitions...
\setlength{\textwidth}{6.5in}
\setlength{\textheight}{9.0in}
\setlength{\oddsidemargin}{0pt}
\setlength{\evensidemargin}{\oddsidemargin}
\setlength{\topmargin}{-0.45in}
\setlength{\headheight}{0.55in}
\setlength{\headsep}{0.10in}
\setlength{\topskip}{12pt}
\setlength{\footskip}{14pt}
\setlength{\parindent}{1em}
%--- End of page format.

%--- fancy page headers on following pages...
\setlength{\headrulewidth}{1pt}
\lhead{\large\slshape Stellar populations and ISM out to the Virgo Cluster\\[-6pt]
{\footnotesize\upshape Science White Paper for the Astro2010 Decadal Survey}\\[-14pt]}
\chead{}
\rhead{\large\upshape R.A.\ Jansen et al.\\[-6pt]
{\footnotesize Page \arabic{page}}\\[-14pt]}
%--- (not so) fancy page footers...
\cfoot{\color{blue}\rule{\txw}{0.1pt}\\[-0pt]\hspace*{-2.3em}\scriptsize\textsl{
A Systematic Study of the Stellar Populations and ISM in Galaxies out to the
Virgo Cluster: near-field cosmology within the local universe}}
\pagestyle{fancy}

\null

\noindent\hspace*{0.5in}\begin{minipage}[t]{6.0in}{\small 
{\bf Abstract:} \ We present a compelling case for a \emph{systematic} and
\emph{comprehensive} study of the resolved and unresolved stellar
populations, ISM, and immediate environments of galaxies throughout the
local volume, defined here as $D\!<\!20$\,Mpc.  This volume is our cosmic
backyard and the smallest volume that encompasses environments as different
as the Virgo, Ursa Major, Fornax and (perhaps) Eridanus clusters of
galaxies, a large number and variety of galaxy groups (e.g., Sculptor,
M\,81, M\,83, CVn\,I and II clouds, M\,51, M\,101, M\,74, NGC\,5866, M\,104,
and M\,77 groups), and several cosmic void regions.  In each galaxy, through
a pan-chromatic ($\sim$160--1100\,nm) set of broad-band and diagnostic
narrow-band filters, ISM structures and individual luminous stars to
$\gtrsim$1\,mag below the TRGB should be resolved on scales of $<$5\,~pc (at
$D\lesssim20$\,Mpc, $\lambda$$\sim$800\,nm, for
$\mu_I\gtrsim24$\,mag\,arcsec$^{-2}$ and $m_{I}^{\hbox{\tiny
TRBG}}\lesssim27.5$\,mag).  Resolved and unresolved stellar populations
would be analyzed through color-magnitude and color-color diagram fitting
and population synthesis modeling of multi-band colors and would yield
physical properties such as spatially resolved star formation histories. 
The ISM within and around each galaxy would be analyzed using key
narrow-band filters that distinguish photospheric from shock heating and
provide information on the metallicity of the gas.  Such a study would
finally allow unraveling the global and spatially resolved star formation
histories of galaxies, their assembly, satellite systems, and the
dependences thereof on local and global environment within a truly
representative cosmic volume. 
The proposed study is not feasible with current instrumentation but
argues for a wide-field ($\gtrsim$250 arcmin$^2$), high-resolution
($\lesssim$0\farcs02--0\farcs065\,[300--1000\,nm]), 
ultraviolet--near-infrared imaging facility on a 4\,m-class space-based
observatory.\\

} \end{minipage}

\bigskip

\noindent{\huge\bf A}t the present epoch, most of the baryonic matter that
condensed into galaxies is locked into stars.  The stellar populations of
galaxies not only record the history of baryonic matter, \eg, through
chemical abundances and stellar spatial distributions, but also its rate of
evolution via the star formation process.  The visible forms of galaxies are
shaped by a series of complex processes which convert dissipative
interstellar matter into nearly collisionless stars.  Despite the success of
current theoretical models in following the growth of dark matter
structures, significant problems remain in understanding how the baryonic
components of galaxies develop.  These include, for example, the low numbers
of visible dwarfs and satellite galaxies relative to the predicted swarms of
low mass dark matter halos around giant systems and the comparatively high
angular momenta and old ages of galactic disks.  Whether these difficulties
represent fundamental issues with the hierarchical dark matter model, a
lack of understanding of star formation processes and their feedback on
galactic scales, or a lack of representative data spanning the full range in
cosmic environment is yet unclear. 

\bigskip

\noindent To advance our understanding of the star formation and chemical
enrichment histories of the stellar systems within the $D<20$\,Mpc local
volume, one would need access to the vacuum UV through near-IR wavelength
regime.  
The UV is uniquely sensitive to hot sources such as massive young stars,
low-mass accreting protostars, and certain types of old, highly evolved
stars.  
Deep UV observations shortward of 365\,nm of A and F-type stars, for example,
are particularly important for tracking metal enrichment, star formation
histories, and galaxy disk evolution. In older ($>$5\,Gyr)\linebreak

  %*%----- footnote does not seem to work on 1st page ----------------------
\smallskip
\noindent{\color{blue}\rule{\txw}{0.1pt}\\[-4pt]\scriptsize\textsl{
A Systematic Study of the Stellar Populations and ISM in Galaxies out to the
Virgo Cluster: near-field cosmology within the local universe}}
  %*%-----------------------------------------------------------------------
\clearpage

\noindent stellar populations, helium-burning stars in advanced evolutionary
phases have surface temperatures $>$10,000\,K, making them UV-bright.  These
hot objects are not only important in their own right, but also provide key
information on mass loss during the red giant branch (RGB) evolution which
precedes the hot phases.  Stellar mass loss is a central problem in stellar
astrophysics and is related to a number of other important processes, such
as dust production, X-ray emission, and accretion flows.  Many key
diagnostics of interstellar gas and dust (ISM) are found also only at
wavelengths shortward of 400\,nm, including the 217.5\,nm peak in the dust
extinction law, and a number of important plasma emission lines (\eg,
\OII\lamlam372.7nm, \ion{Mg}{2}\,\lam279.9nm and \Lya).  The 150--250\,nm
region is also one of the \emph{darkest} parts of the natural sky background
above the Earth's atmosphere, permitting the detection of extremely faint
sources.  \emph{Wide-field, high-resolution vacuum-UV imaging would open up
a new window on this last under-explored corner of normal stellar evolution}.

\medskip

\noindent Stellar populations contain the histories of evolution of the
baryonic components of galaxies.  Accessing this information is complicated
by the presence of multiple stellar population components projected along
each sightline, effects of interstellar dust on observed spectral energy
distributions, and the relatively low brightnesses of outer regions of
galaxies relative to the sky.  Multi-band UV through near-IR
($\sim$200--1100\,nm) measurements from space are required to derive
extinction corrected stellar SEDs with sufficient precision to distinguish
differences in metallicity and age.  Unraveling the star formation histories
of nearby galaxies (and spatial variations therein) \emph{in detail}
requires one to resolve individual stars to $\gtrsim$1\,mag below the Tip of
the RGB (TRGB).  Although \HST\ would in theory be capable of accessing
the TRGB out to $\sim$12\,Mpc, in practice \emph{very} few studies have been
able to push beyond 7\,Mpc because of \HST's limited aperture
(exposure times comparable to those in the Deep and Ultradeep Fields would
be required).  A particularly novel opportunity enabled by a larger aperture
and similar or higher resolution would be to resolve red K-giant stars
within star streams known to exist within the Virgo Cluster in the form of
structure in the diffuse intra-cluster light.  This would allow unraveling
for the first time the 3D structure, kinematics and galaxy assembly history
within the Virgo Cluster.  \emph{Space-based wide-field high-resolution
optical--near-IR imaging would open up a new window of discovery space that
remains inaccessible to or exceedingly inefficient with next-generation
giant ground-based telescopes}. 

\medskip

\noindent Previous space-based UV--near-IR imaging facilities, however,
emphasized either low spatial resolution and wide fields (\eg, \GALEX;
strictly UV, minimal filter set [$n$=2]) or high resolution and small fields
(\eg, \HST).  For a study of both the resolved stellar populations
\emph{and} its dependence on the global structure and evolution of nearby
galaxies, one would need to combine:\\[-24pt]

\begin{itemize}
\setlength{\itemsep}{-4pt}
\setlength{\itemindent}{-5.5ex}
\setlength{\labelwidth}{-5.5ex}

\item[\textsl{\bfseries(1)}] a large field of view (FoV) that is well-matched
to the angular sizes of nearby galaxies and their satellite systems;

\item[\textsl{\bfseries(2)}] sensitivity to detect at $\gtrsim$\,5$\sigma$
individual RGB stars to $\gtrsim$1\,mag below the Tip of the RGB
(TRGB; $M_{I}^{\hbox{\tiny TRBG}}$$\simeq$$-$4.0\,mag, 
$m_{I}^{\hbox{\tiny TRBG}}\lesssim$27.5\,mag);

\item[\textsl{\bfseries(3)}] high angular resolution that allows resolving
individual luminous RGB stars at linear scales of $<$\,5\,pc out to
$\sim$20\,Mpc (at $\lambda$$\sim$800\,nm and
$\mu_I\gtrsim24$\,mag\,arcsec$^{-2}$); and

\item[\textsl{\bfseries(4)}] a sufficiently rich complement of UV--near-IR 
broad-, medium- and narrow-band filters to provide physically meaningful
diagnostics on both stars and ISM. 
\end{itemize}

\clearpage

%%%%%%%%%%%%%%%%%%%%%%%%%%%%%%%%%%%%%%%%%%%%%%%%%%%%%%%%%%%%%%%%%%%%%%%%%%%%
%
\color{blue}
\noindent\leavevmode
\framebox[\textwidth]{
   \begin{minipage}[t][0.450\textheight][t]{\textwidth}
   \centering
   \includegraphics[width=0.99\textwidth]{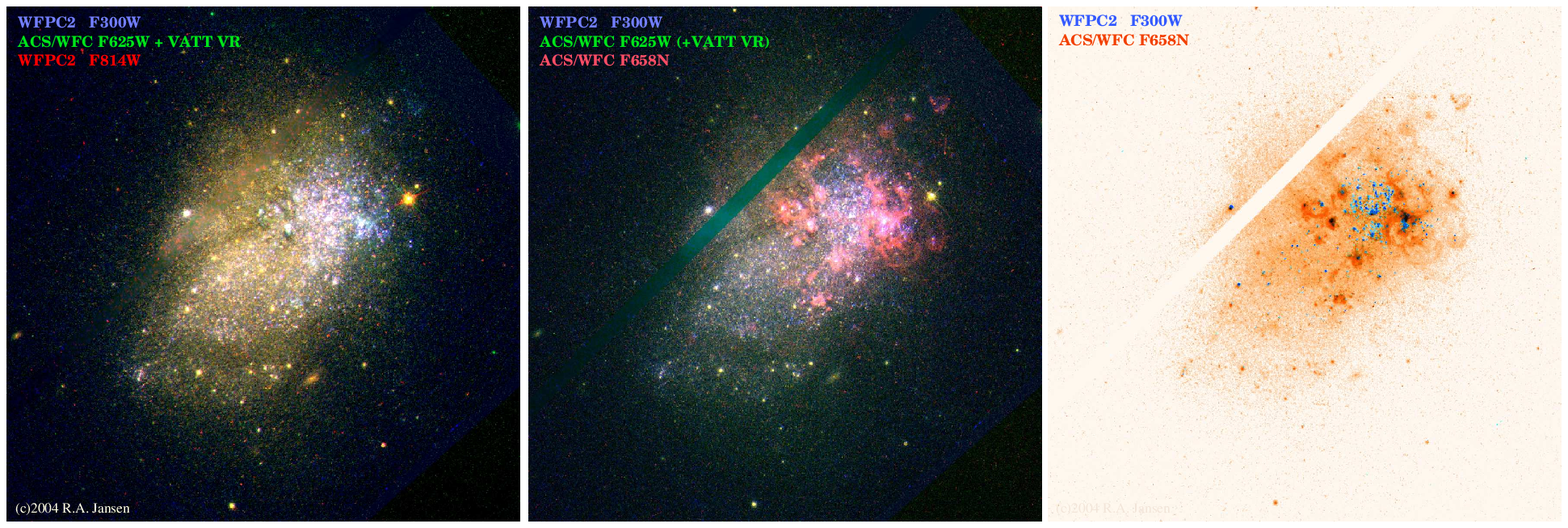}
   \parbox[t]{0.985\textwidth}{\vspace*{-12pt}\color{black}\small Fig.~1 ---
Three views of NGC\,3738, a nearby Irregular galaxy, whose star formation
history is characterized by episodes of vigorous star formation.  (\emph{a})
broad-band filters highlight the spatial distribution of stellar populations
of various ages, (\emph{b}) broad- and narrow-band images highlight the
interplay between star formation and the ISM, the deposition of mechanical
energy, (\emph{c}) a mid-UV--\Ha\ composite highlights the relation
between the hot young stars that dominate the mid-UV and the ionized ISM. 
Various stages of the star-formation cycle, from deeply embedded and
obscured star formation to cluster formation and break-out, can be
identified.  Whereas \HST\ observations like these tend to be shallow and
rarely provide simultaneous full coverage and $<$5\,pc resolution, we here
advocate a systematic study of the star formation histories and current
massive star formation and its feedback within galaxies and their
surrounding satellite galaxy systems, in cosmic environments as different as
rich clusters, galaxy groups and void regions.}
   \end{minipage}
}
\color{black}
%
%%%%%%%%%%%%%%%%%%%%%%%%%%%%%%%%%%%%%%%%%%%%%%%%%%%%%%%%%%%%%%%%%%%%%%%%%%%%

\smallskip

%%%%%%%%%%%%%%%%%%%%%%%%%%%%%%%%%%%%%%%%%%%%%%%%%%%%%%%%%%%%%%%%%%%%%%%%%%%%
%
\begin{wrapfigure}[23]{r}{0.50\textwidth}
   \color{blue}\vspace*{-18pt}\framebox[0.500\textwidth]{
      \begin{minipage}[t][0.520\textheight][t]{0.485\textwidth}
         \centering
         \includegraphics[width=0.99\textwidth]{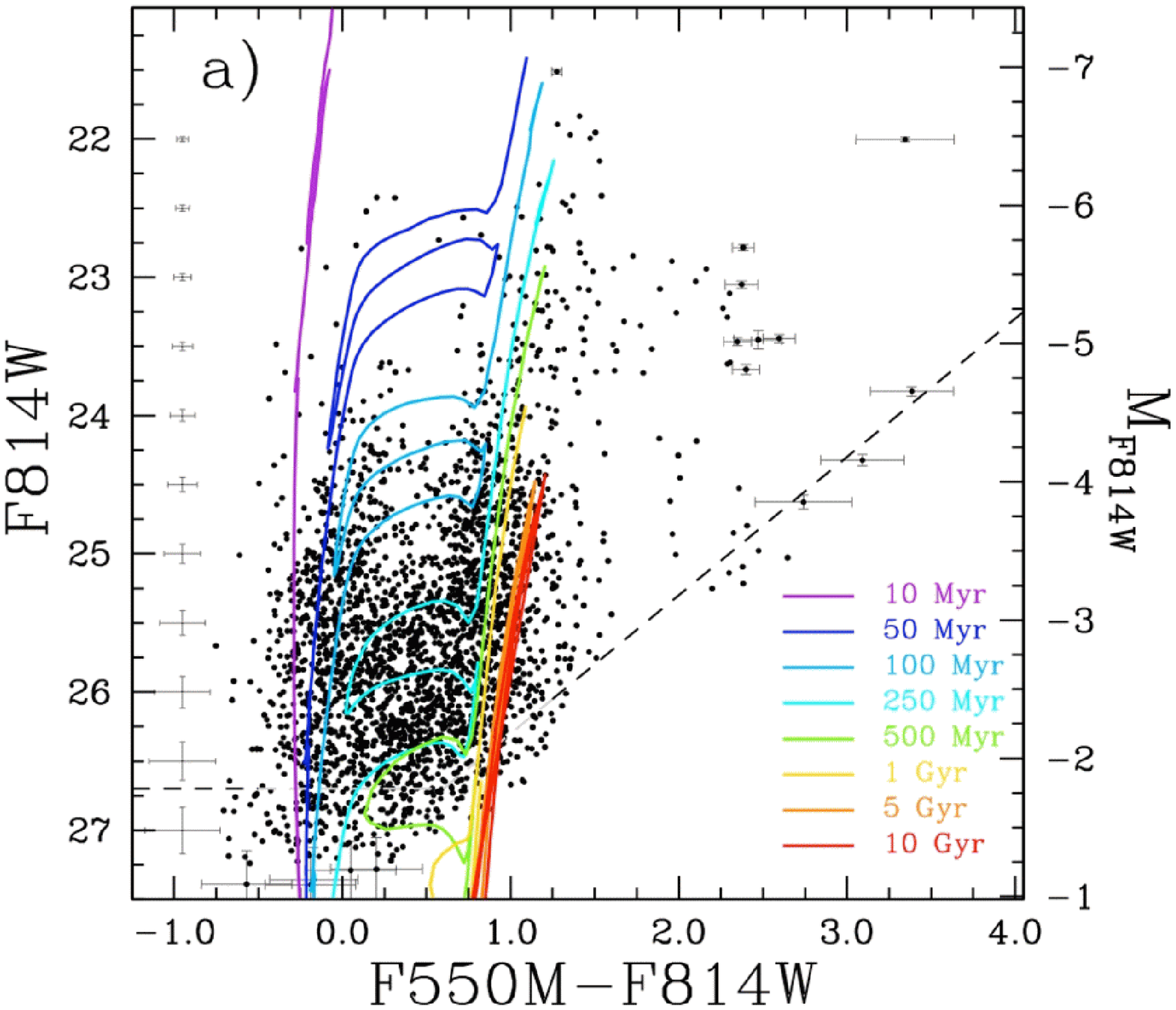}  
         \parbox[t]{\textwidth}{\vspace*{-8pt}\color{black}\small
Fig.~2 --- \HST$/$ACS-HRC $I$ vs.\ ($V\!\!-\!\!I$) color-magnitude diagram
of the resolved stellar populations within nearby metal-poor dwarf irregular
galaxy CGCG\,269-049 (Corbin \etal 2008).  Padua isochrones are overlayed,
demonstrating a complex star formation history with stellar populations of
ages ranging from $<$100\,Myr to $\gtrsim$10\,Gyr.  At 5\,Mpc, the TRGB is
discernable at $I$\,=\,24.5\,mag.  We advocate a UV--near-IR facility
that would deliver CMDs of similar quality out to the Virgo Cluster.}
      \end{minipage}
   }\color{black}
\end{wrapfigure}
%
%%%%%%%%%%%%%%%%%%%%%%%%%%%%%%%%%%%%%%%%%%%%%%%%%%%%%%%%%%%%%%%%%%%%%%%%%%%%

\noindent Color-magnitude and color-color diagrams obtained by \HST\ and
large ground-based telescopes of the resolved stellar populations within
nearby galaxies enabled enormous leaps forward in our understanding of the
stellar mass distributions and star-formation histories within our own and
nearby ($D\,\lesssim\,$6\,Mpc) galaxies.  Pushing that capability out to
20\,Mpc, with large fields of view sampled with sufficient sensitivity and
sampled at linear scales $<$5\,pc would provide access to galaxies within a
large and varied number of galaxy groups as well as the nearest clusters
(Ursa Major and Virgo). 

\medskip

\noindent For higher-surface brightness regions and for serendipitously
observed more distant galaxies, where individual stars cannot be resolved,
the constituent stellar populations can still be deduced from their
integrated light, since the integrated UV energy distributions of stellar
populations evolve strongly over timescales up to 3\,Gyr.  The
high-sensitivity region is shortward of \tsim400\,nm, where the confluence
of hydrogen absorption lines in hotter stars and the Balmer Jump and
metallic absorption features in cooler ones begin to strongly affect the
gross spectral structure.  The UV will provide the parameters needed to
measure star formation rates and break age--metallicity degeneracies in
disecting composite stellar populations, and recover star-formation
histories. 

\noindent The UV allows direct detection of the massive stars responsible
for most of the ionization, photo-dissociation, kinetic-energy input, and
element synthesis in galaxies.  These processes are responsible for much of
the astrophysics of the universe.  By contrast, most other methods of
studying massive star populations yield only indirect measures since they
rely on re-processing of the UV photons by the surrounding medium (\HII\
regions or dust clouds).  Furthermore, since the production of
Lyman-continuum photons by young populations rapidly declines after
\tsim5\,Myr, these other methods probe star formation only over a short
period, which constitutes a tiny fraction (0.05\%) of the lifetime of a
galaxy.  By comparison, the short-wavelength continuum below 400\,nm remains
a sensitive indicator of star-formation histories for ages up to 100$\times$
greater. 

\bigskip

\noindent {\bfseries Key scientific themes that have arisen from recent
advances}

\noindent {\bfseries\itshape Near Field Cosmology: the oldest stellar
populations.} \ It is useful to ask \emph{where} the oldest stars are
located.  We know that in the Milky Way they reside in the spheroidal halo,
in the LMC and SMC they have the largest radial scale of any stellar
population, and they usually are the least centrally concentrated stellar
component of dwarf spheroidals.  However, even in nearby galaxies these
results only apply in a mean sense.  With the growing realization of the
importance of interactions in the lives of galaxies, as demonstrated by the
discovery of tidal debris streams and plumes in, e.g., the Milky Way, M\,31,
M\,81, and NGC\,4013, the old star distribution merits reexamination.  Are
older stars asymmetrically distributed in the outer regions of galaxies, as
expected if they were contributed by dissolving satellites? Data for
inclined galaxies also will provide information on globular cluster systems,
bulge vs.\ disk stellar populations, disk vertical structures, dust lane
forms, warps, and a variety of other features on intermediate galactic
scales.  Each of these connects in useful ways to the evolutionary history
and thus provides an empirical base for application of the expertise of the
astronomical community. 

\smallskip

\noindent {\bfseries\itshape Star Formation and its Products.} \ The
existing combination of ground-based and \HST\ imaging provides an excellent
base from which to design investigations of the nature and extent of star
forming sites.  Investigations of connections between drivers, if any, for
star formation --- spiral arms, interactions, etc., as well as basic
galactic properties --- are essential for understanding how feedback
operates.  A survey of the local $D$$<$\,20\,Mpc volume provides the range of
galaxy types, luminosities, cosmic environment, and the sensitivity and
statistics to support a complete study of the association of compact
clusters and regions of star formation.  From programs like \SINGS\ and
other recent and ongoing ground-based surveys, global star formation rates
and time scales are anticipated to be known.  We then can compare the small
scale characteristics of star formation, an intrinsically local process,
with the overall galactic environment.  Are these statistically connected
and, if so, how?  By combining deep mid-UV and narrow-band \Ha\ observations,
it becomes possible to also address the escape fraction of ionizing
radiation in a variety of galaxies.  This question is particularly important
in low metallicity dwarf galaxies which may have traits in common with the
types of objects responsible for finishing reionization of the universe at
redshifts $z>$\,6. 

\smallskip

\noindent {\bfseries\itshape Are Galactic Disks Growing?} \ As already
demonstrated by \GALEX, young stars have high contrast against the sky in
the mid-UV.  This spectral range therefore opens the way for mapping star
formation in low-density environments, including the outer disks of
galaxies.  A next-generation wide-field UV--near-IR space observatory must
offer major advantages in sensitivity and resolution over the pioneering
results from \GALEX.  Hence, we would be able to determine ages and
photometric stellar masses for small star forming complexes of the type that
appear to populate the outer disks of galaxies, ranging from small
irregulars to giant spirals.  From these, star formation rates per unit area
and, thus, disk growth rates can be estimated. 

\smallskip

\noindent {\bfseries\itshape Galactic Centers.} \ Centers of galaxies are
dumping grounds.  Baryonic material that ends up in the central zone of a
galaxy has experienced substantial dissipation and loss of angular momentum. 
Yet it is not uncommon to find high-density stellar and gas systems
coexisting within 1\,kpc of the centers of galaxies.  Centered in this zone
are the nuclei themselves, many of which harbor massive black holes.  We
would be able to systematically chart the stellar properties of nuclear
environments.  Where and in what ways are stars formed (clusters, scaled
OB~associations, spiral arms, rings, clumps)? How does star formation relate
to the properties of nuclei on small scales and on the other side to the
surrounding main disk? How are bars, both large and nuclear, related to the
structure and activity levels in nuclei?

\smallskip

\noindent {\bfseries\itshape A Survey of Nearby Galaxies.} \ We propose to
learn how galaxies work, through studies of their stars, ISM, and immediate
environments, and to build the definitive UV--near-IR photometric imaging
database of galaxies within our local slice of the Universe.  This would
result in a 21st century digital `Hubble Atlas' of nearby galaxies and their
surroundings that will provide a standard for testing our understanding of
how galaxies attained their present forms and how their stellar components
will likely evolve into the future.  The resolved and unresolved stellar
populations would be analyzed through color-magnitude and color-color
diagram fitting, providing accurate and uniform TRGB distances, and through
population synthesis modeling of multi-filter broad- and medium-band
photometry.  The ISM in each galaxy would be observed through key
narrow-band filters (\Ha, \Hb, \OII, \OIII, \SII; possibly \ion{Mg}{2}, \OI,
\ion{Ca}{2} or \SIII) that allow identifying the ionized gas, estimate
its metallicity and variations therein, and for each region determine
whether ionization is dominated by photospheric or shock heating.  Direct
measurements of the extinction toward that ionized gas through the Balmer
decrement (\Ha/\Hb) will allow measuring the currently ongoing, high-mass
star formation in each star formation region. 

\bigskip

\noindent{\bfseries Key advances in observation needed}

\noindent \emph{Resolution} --- $\lesssim$0\farcs02--0\farcs065 [300--1000\,nm]
resolution is required in order to resolve luminous RGB stars out to the
distance of the Virgo Cluster, and to resolve the relevant scales for star
formation feedback processes within the ISM (shocks, outflows, bubbles,
shells) within galaxies.

\noindent \emph{Wavelength agility} --- access to both vacuum-UV and near-IR;
no wavelength regime alone will suffice for a comprehensive understanding of
the star-formation and assembly histories of galaxies.

\noindent \emph{Wide-field focal plane arrays} --- these are presently not
at sufficiently high TRL; investment is needed to improve yields, provide
cheaper devices and high-throughput assembly and testing to enable economies
of scale.  Such an investment would not just benefit the science proposed here. 

\noindent \emph{Coatings} --- an investment in improving the relatively poor
broad-band performance of optical coatings of telescope mirrors in the UV,
with typical reflectances below 85\% (Al+MgF$_2$) and 65\% (Al+LiF), directly
results in a large increase in throughput for a given telescope aperture, or
more affordable missions for a given sensitivity requirement.

\noindent \emph{Dichroics} --- most photons collected by telescopes are
rejected by bandpass filters.  Dichroic(s) potentially double (or even triple)
the observing efficiency of astronomical observatories (\eg,
\Spitzer/IRAC) and allow tuning downstream optics and detectors for
more optimal performance, avoiding compromises inherent in forcing performance
over more than an octave in frequency.

\bigskip

\noindent{\bfseries Enabling science investigations}

\noindent The proposed science in the present white paper does not stand
alone, but must build on a strong understanding of the physics of the
star formation process in various environments.  Gaining such
understanding requires observational detail that can only be attained
within our own Galaxy.  From that basis one has to step out to galaxies
spanning a range of metallicity and star formation activity within our
Local Group to provide observational tracers with calibrations that are
directly and solidly rooted in physics.  We refer the reader to the
Science White Paper by P.~Scowen \etal ``Understanding Global Galactic
Star Formation''.  Also, investment in human capital and in ground-based
supporting and path-finding programs, including operational support,
should not be ignored, as the overal science return of this and many
`high-end' programs critically depends on it. 

\bigskip

\noindent{\bfseries Four central questions to be addressed}

\vspace*{-8pt}\noindent\begin{itemize}
\setlength{\itemsep}{-4pt}
\setlength{\itemindent}{-5.5ex}
\setlength{\labelwidth}{-5.5ex}

\item[\textsl{\bfseries(1)}] What is the spatially resolved star formation
history of a comprehensive and representative subset of the galaxies
encompassed within the local $D$$<$\,20\,Mpc volume? To what extent and how
does it depend on morphological type class, mass, and cosmic environment?
Does this fossil record confirm in detail the broad picture inferred from
the evolution of the cosmic star formation density?  What does it tell us
about the formation and survival rates of solar systems like our own in
different galaxies?

\item[\textsl{\bfseries(2)}] What is the mass assembly history of galaxies
within the varied cosmic environments encompassed within the local
$D$$<$\,20\,Mpc volume, the smallest representative slice of the Universe?
This overarcing question will likely also include the questions: do galaxies
grow from the inside out or is this too simple a picture; and why do galaxy
disks have such high angular momenta; and why are they so old?

\item[\textsl{\bfseries(3)}] Can we unravel the true 3-D structure and
internal kinematics of galaxy groups and the Ursa Major and Virgo Clusters
via reliable TRGB distances and fossil star streams/intra-cluster light? Can
we meaningfully constrain how baryons found their way from the IGM into the
galaxies, galaxy groups and into clusters of galaxies within the local
Universe?

\item[\textsl{\bfseries(4)}] Why does there at least seem to be a dearth of
satellite galaxies around the primary galaxies within our Local Group
compared to predictions from $\Lambda$CDM numerical simulations? Is the
Local Group result confirmed throughout the local volume, and if so, what
fundamental factor is missing in the simulations?
\end{itemize}

\bigskip

\noindent{\bfseries Area of unusual discovery potential for the next decade}

\noindent Combination of a large collecting area, very wide field of view,
high angular resolution, wavelength agility and/or multiplexing advantage
would allow orders of magnitude more efficient UV--optical observations of
the star formation and many other processes and, moreover, open up a new
domain in discovery space near and far. 
  %
  %*%\noindent 
  %
Injection into L2 (or Earth Drift-Away) orbits allows provide dynamical and
thermal stability, and increases (doubling) in efficiency over LEO orbits
and, hence, lower cost per hour of observation (all other variables being
equal).
  %
  %*%\noindent
  %
Large focal plane array (douzens to hundreds of individual CCD or CMOS
detectors) and dichroic camera (simultaneous observation in two or more
channels of the same field of view) technology is better matched to the
collimated beams provided by optical telescope assemblies and less
wasteful in terms of collected photons, maximizing science output and
especially benefitting survey science with a lasting legacy beyond the
nominal duration of a mission.
  %
  %*%\noindent
  %
Survey science allows discovery of very rare objects amongs billions and
billions, the positions an properties of which may not be knowable a priori.

\vspace*{0.5in}

{\small
\noindent{\bfseries References}\\[-1pt]
Corbin, M., Kim, H., Jansen, R., Windhorst, R., \& Cid\,Fernandes, R. 2008,
	ApJ 675, 194\\[-2pt]
Scowen, P., Jansen, R., Beasley, M., \etal 2009, \emph{``Understanding Global
	Galactic Star Formation''}, Science\\[-2pt]
	$\null\qquad$White Paper submitted to the \emph{Astro2010} Decadal
	Survey, Feb 15.
}

\end{document}